\documentclass[conference]{IEEEtran}
\IEEEoverridecommandlockouts

\usepackage{cite}
\usepackage{amsmath,amssymb,amsfonts}
\usepackage{algorithmic}
\usepackage{graphicx}
\usepackage{textcomp}
\usepackage{xcolor}
\usepackage{subcaption}
\usepackage{tikz}
\usetikzlibrary{patterns}
\usepackage{makecell} 
\definecolor{stableIP}{RGB}{32,178,170}
\definecolor{IPChange}{RGB}{178,32,39}
\definecolor{H3}{RGB}{111,45,168}
\definecolor{H2}{RGB}{103,168,45}

\def\BibTeX{{\rm B\kern-.05em{\sc i\kern-.025em b}\kern-.08em
    T\kern-.1667em\lower.7ex\hbox{E}\kern-.125emX}}
\begin{document}

\title{Performance Comparison of HTTP/3 and HTTP/2 with Proxy Integration}

\author{
\IEEEauthorblockN{Fan Liu, Behrooz Farkiani, John Dehart, Jyoti Parwatikar, Patrick Crowley}
\IEEEauthorblockA{\textit{Washington University in St. Louis}, St. Louis, MO, USA \\ 
\{fan.liu, b.farkiani, jdd, jp, pcrowley\}@wustl.edu}
}

\maketitle

\begin{abstract}
This paper systematically evaluates the performance of QUIC/HTTP3 (H3) and TCP/HTTP2 (H2) in proxy-enhanced environments. H3 integrates UDP-based flow-controlled streams, built-in TLS, multiplexing, and connection migration to better support modern web communication. While prior studies show that H3 can outperform or underperform H2 depending on network conditions, the role of proxies and connection migration remains underexplored. We assess a variety of H2 and H3 client implementations across, particularly in lossy networks and proxy setups. Our findings show that proxies can significantly enhance H2 performance, yielding a 90\% improvement in single-stream downloads under severe impairments when used with the BBR congestion control algorithm. In contrast, proxies have minimal impact on H3, which maintains consistent performance due to its internal mechanisms. H3 excels under high-loss and high-latency conditions, leveraging connection migration and multiplexing to deliver up to 88.36\% improvement in migration scenarios and 81.5\% in extreme loss cases. While optimized H2 can match H3 in some settings, H3 is more robust overall, showing less sensitivity to proxies, impairments, and congestion control variations.
\end{abstract}


\section{Introduction}
The rapid evolution of network technologies has driven advancements in transport and application-layer protocols, notably QUIC \cite{quicietc_2022} and HTTP/3 \cite{http3ietc_2022}. Developed by Google and later standardized by the IETF, QUIC underpins HTTP/3, introducing faster connection establishment, multiplexing without head-of-line (HoL) blocking, and seamless connection migration across changing networks \cite{quicietc_2022}. HTTP/3 leverages these features to improve web performance through efficient packet handling and prioritization \cite{http3ietc_2022}.

Although QUIC and HTTP/3 have been widely studied \cite{Jaeger2023QUICOT, Liu2023SecurityAP, Carlucci2015HTTPOU}, their behavior in proxy-enhanced environments remains underexplored. Proxies are integral to modern web architectures, supporting content delivery and load balancing. However, the impact of proxies on H3's performance, especially under connection migration and network impairments, remains unclear.

One challenge lies in how QUIC’s encryption and connection-oriented nature interact with proxies, which terminate and reinitiate connections. This could introduce overhead or degrade performance. While QUIC supports connection migration, it is uncertain how proxies influence its effectiveness in real-world scenarios. Understanding these interactions is essential for optimizing deployments in proxy-heavy environments such as mobile networks, CDNs, and enterprise systems. Moreover, empirical performance data from such evaluations can inform machine learning models for automated protocol selection and network optimization in modern data-driven network management systems.

This study addresses these gaps by systematically comparing the performance of TCP-based HTTP/2 (H2) and QUIC-based HTTP/3 (H3) under both proxy and non-proxy conditions, focusing on realistic network scenarios. We measure key metrics such as request time and Speed Index (SI) across single-stream, multi-stream, connection migration, and webpage loading experiments.

Specifically, we address the following questions:
\begin{itemize}
    \item How does QUIC’s connection migration perform with and without proxies?
    \item Under what conditions do proxies enhance or degrade H2 and H3 performance?
    \item How do congestion control algorithms affect protocol performance?
    \item Do proxies introduce overhead to H2 and H3?
\end{itemize}

Our contributions include:
\begin{itemize}
    \item \textbf{Performance Evaluation with Proxy Integration:} We show that proxies significantly improve H2 performance with BBR, while H3 maintains consistent behavior due to its built-in mechanisms.
    \item \textbf{Connection Migration Analysis:} We assess QUIC’s migration performance in proxy scenarios, highlighting H3’s resilience during network transitions.
    \item \textbf{Congestion Control Impact:} We analyze how different CCAs influence H2 and H3, showing H3’s robustness across algorithms.
\end{itemize}

\begin{table*}[t]
\centering
\caption{Studies of QUIC Connection Migration Under Proxy Environments}
\begin{tabular}{|p{2.3cm}|p{2.2cm}|p{2.5cm}|p{3.2cm}|p{2.7cm}|p{2.5cm}|}
\hline
\textbf{Study} & 
\textbf{Protocol} & 
\textbf{Proxy Integration} & 
\textbf{Connection Migration} & 
\textbf{Multi-client Testing} & 
\textbf{Network Conditions} \\
\hline

Kosek et al. \cite{Kosek2022ExploringPQ} & H1.1(TCP); H3(QUIC) & PEP over satellite networks & Not studied & Single client & Varied one-way loss and delay \\ \hline

Border et al. \cite{border2020evaluating} & H1.1(TCP/gQUIC); H3(TCP/gQUIC) & PEP over satellite networks & Not studied & Single client & Varied bandwidth, loss and delay \\ \hline

Kosek et al. \cite{kosek2023secure} & SMAQ & PEP over satellite networks & Used in their SMAQ design, but didn't show the performance of this feature & Single client & Varied one-way loss and delay \\ \hline

Kühlewind et al. \cite{kuhlewind2021evaluation} & QUIC Only & MASQUE proxying & Not studied & Single client & Fixed one-way delay and varied loss \\ \hline

\textbf{Our Work} & \textbf{H2(TCP); H3(QUIC)} & \textbf{Reverse proxy} & \textbf{Detailed evaluation with proxy integration} & \textbf{6 clients across different implementations} & \textbf{Varied two-way loss and delay} \\
\hline
\end{tabular}
\label{tab:connection_migration}
\end{table*}

\section{Related Work}
QUIC and HTTP/3 have been widely studied for their advantages over TCP-based protocols. However, their combined behavior with proxies and during connection migration remains underexplored.

\textbf{General QUIC and HTTP/3 Performance}  
Comparative studies show that H3 can outperform or underperform H2 depending on conditions \cite{Cunha2023}. While H3 avoids HoL blocking and supports migration, its performance varies with prioritization strategies and implementation maturity \cite{Marx2019ResourceMA, Sander2022AnalyzingTI}. In high-speed environments, H2 sometimes outperforms H3 due to QUIC’s overhead \cite{Zhang2024QUIC}. QUIC has also shown advantages in web and video traffic \cite{Shreedhar2022EvaluatingQP, Wolsing2019APP}, and has been extended through MPQUIC \cite{Coninck2017MultipathQD} and optimized using advanced scheduling \cite{Paiva2023AFL}.

\textbf{QUIC Connection Migration}  
QUIC’s connection migration supports session continuity across network changes. Although a Google demo showcases its benefits, empirical research is limited. \cite{Tan2020ProactiveCM} proposed proactive migration strategies, but proxy interactions remain unstudied.

\textbf{QUIC and Proxies}  
Few studies examine QUIC with proxies. Secure Middlebox-Assisted QUIC (SMAQ) \cite{kosek2023secure} and PEP-based enhancements \cite{Kosek2022ExploringPQ} show performance gains in satellite networks but do not address connection migration. \cite{kuhlewind2021evaluation} explored MASQUE proxying for UDP tunneling but did not assess performance trade-offs with multiplexing or migration. Table \ref{tab:connection_migration} summarizes existing efforts, showing that proxy effects on migration are largely unexamined.

\textbf{Research Gap and Focus}  
While prior work has explored QUIC under diverse conditions, wireless networks, congestion algorithms, and multipath routing, proxy behavior remains a major gap. This paper addresses this gap by systematically comparing H2 and H3 in proxy-enhanced settings, focusing on connection migration, congestion control, and network impairments.

\begin{table*}[t]
\centering
\caption{Experiment Setup}
\label{tab:baseline_benchmarks}
\begin{tabular}{|l|l|c|l|l|l|}
\hline
\textbf{Scenarios} & \textbf{Clients} & \textbf{Server} & \textbf{Network Conditions} & \textbf{Workloads} & \textbf{Metrics} \\
\hline
Single-Stream Request & 
\makecell[l]{Proxygen (H3); Curl (H2, H3) \\ Chrome (H2, H3); Wget (H2) \\ Quiche (H3); aioquic (H3)} & 
Nginx & 
\makecell[l]{Bandwidth: 10 Mbps \\ Loss: 0, 2, 4\% \\ Delay: 0, 25, 50 ms} & 
\makecell[l]{Small Files: 0KB, 10KB, 100KB\\ Medium Files: 500KB, 1MB, 5MB\\ Large Files: 10MB} & 
Total Request Time \\
\hline
Multi-Stream Request & 
Curl (H2, H3) & 
Nginx & 
\makecell[l]{Bandwidth: 10 Mbps \\ Loss: 0, 2, 4, 8, 12\% \\ Delay: 0 ms} & 
20 Files: 1MB each & 
Total Request Time \\
\hline
Connection Migration & 
\makecell[l]{Curl (H2)\\ Wget (H2)\\ Quiche (H3)} & 
Nginx & 
\makecell[l]{Bandwidth: 10 Mbps \\ Loss: 0, 2, 4\% \\ Delay: 0, 25, 50 ms} & 
One 5MB file & 
Total Request Time \\
\hline
Web Page Loading & 
Chrome (H2, H3) & 
Nginx & 
\makecell[l]{Bandwidth: 10 Mbps \\ Loss: 0, 2, 4\% \\ Delay: 0, 25, 50 ms} & 
18 popular local host websites & 
SI \\
\hline
\end{tabular}
\end{table*}

\section{Experimental Setup}
To ensure a controlled and reproducible environment, we conduct experiments on a remotely accessible hardware testbed \cite{wiseman2008remotely}. The client and proxy run Ubuntu 20.04.6 on Intel Xeon E5520 (16GB RAM), and the server runs Ubuntu 22.04.4 on Intel Core i3-3240 (16GB RAM), connected via 1 Gbps Ethernet. We use \texttt{tc netem} to emulate bandwidth, loss, and delay. A fixed 10 Mbps bandwidth is applied in all tests. Loss and delay are varied to reflect realistic edge conditions. This controlled setup aligns with prior work \cite{Yu2021DissectingPO} and complements bandwidth-focused studies like \cite{Zhang2024QUIC}.

Nginx serves as the web server, and Envoy as the reverse proxy. We tested six clients, Proxygen, Curl, Chrome, Wget, Quiche, and aioquic, chosen for QUIC v1 support, open-source availability, popularity, and implementation diversity.

Each experiment is run 10 times. We measure Total Request Time for single-stream, multi-stream, and connection migration scenarios, and use SI for web page loading to quantify user-visible responsiveness.

\subsection{Baseline Experiment Setup}
\begin{figure}[ht]
  \centering
  \includegraphics[scale=0.5]{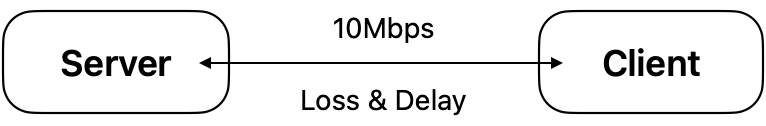}
  \caption{Baseline experiment topology.}
  \label{fig:baseline_topo}
\end{figure}

The baseline setup evaluates H2 and H3 across single-stream, multi-stream, connection migration, and web page loading scenarios. We employ Cubic as the CCA on the client side, and New Reno on the server side for both QUIC and TCP. Cubic is the default CCA for most H3 clients and the default CCA on Linux. For QUIC, the New Reno is the only CCA that Nginx supports. Both H2 and H3 use TLS 1.3. Our experimental setup is detailed in Table \ref{tab:baseline_benchmarks}, with the logical topology illustrated in Figure \ref{fig:baseline_topo}.

\begin{figure}[htbp]
  \centering
  \includegraphics[width=\linewidth]{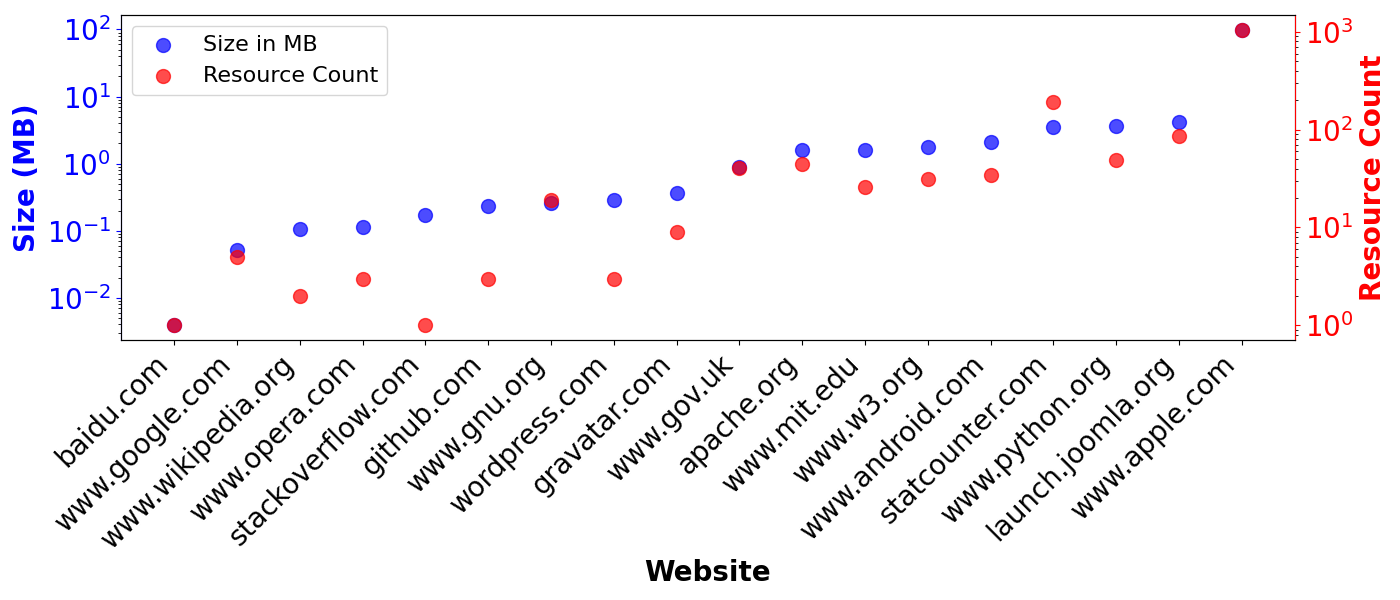}
  \caption{Website sizes and resource counts.}
  \label{fig:websize}
\end{figure}

Workloads include random plaintext files for single/multi-stream scenarios. For web page loading, we follow \cite{Marx2019ResourceMA, Sander2022AnalyzingTI} by selecting 18 popular websites with varying sizes and resource counts, shown in Figure \ref{fig:websize}. Sites are hosted locally on Nginx to eliminate external network effects.

For connection migration, we extend Cloudflare’s quiche client to simulate IP/interface changes, using a design inspired by Android’s Connection Manager. Migration is triggered after 2MB of a 5MB file has been downloaded. For H2, we use wget and curl to test its ability to resume interrupted transfers.

\subsection{Proxy Experiment Setup}
\begin{figure}[ht]
  \centering
  \includegraphics[width=\linewidth]{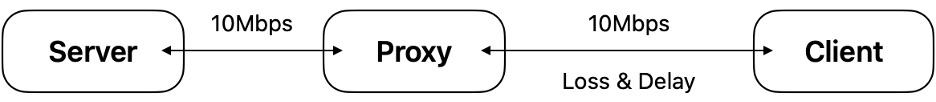}
  \caption{Proxy experiment setup: proxy between client and server.}
  \label{fig:proxy_topo}
\end{figure}

In this setup, Envoy acts as a reverse proxy between the client and server. Loss and delay are applied between client and proxy. The proxy breaks and re-establishes connections, using BBR for both TCP and QUIC, consistent with Envoy defaults. Proxy cache and retry features are disabled. All links are fixed at 10 Mbps. The proxy topology is shown in Figure \ref{fig:proxy_topo}.

\begin{figure*}[ht]
\centering
\begin{subfigure}[b]{0.32\textwidth}
\centering
\includegraphics[width=\linewidth]{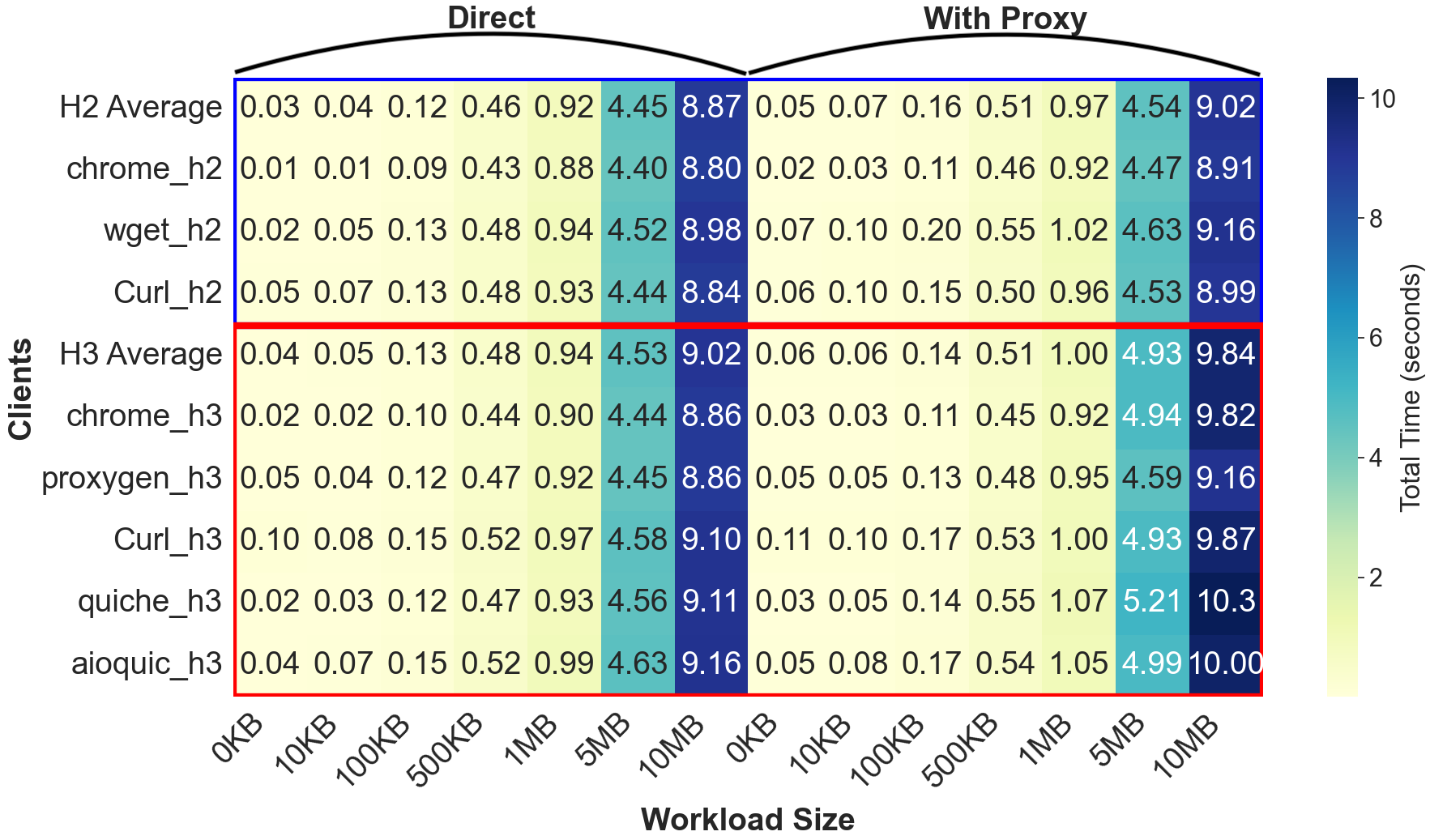}
\caption{0\% Loss, 0ms Delay}
\label{fig:1a-s}
\end{subfigure}
\hfill
\begin{subfigure}[b]{0.32\textwidth}
\centering
\includegraphics[width=\linewidth]{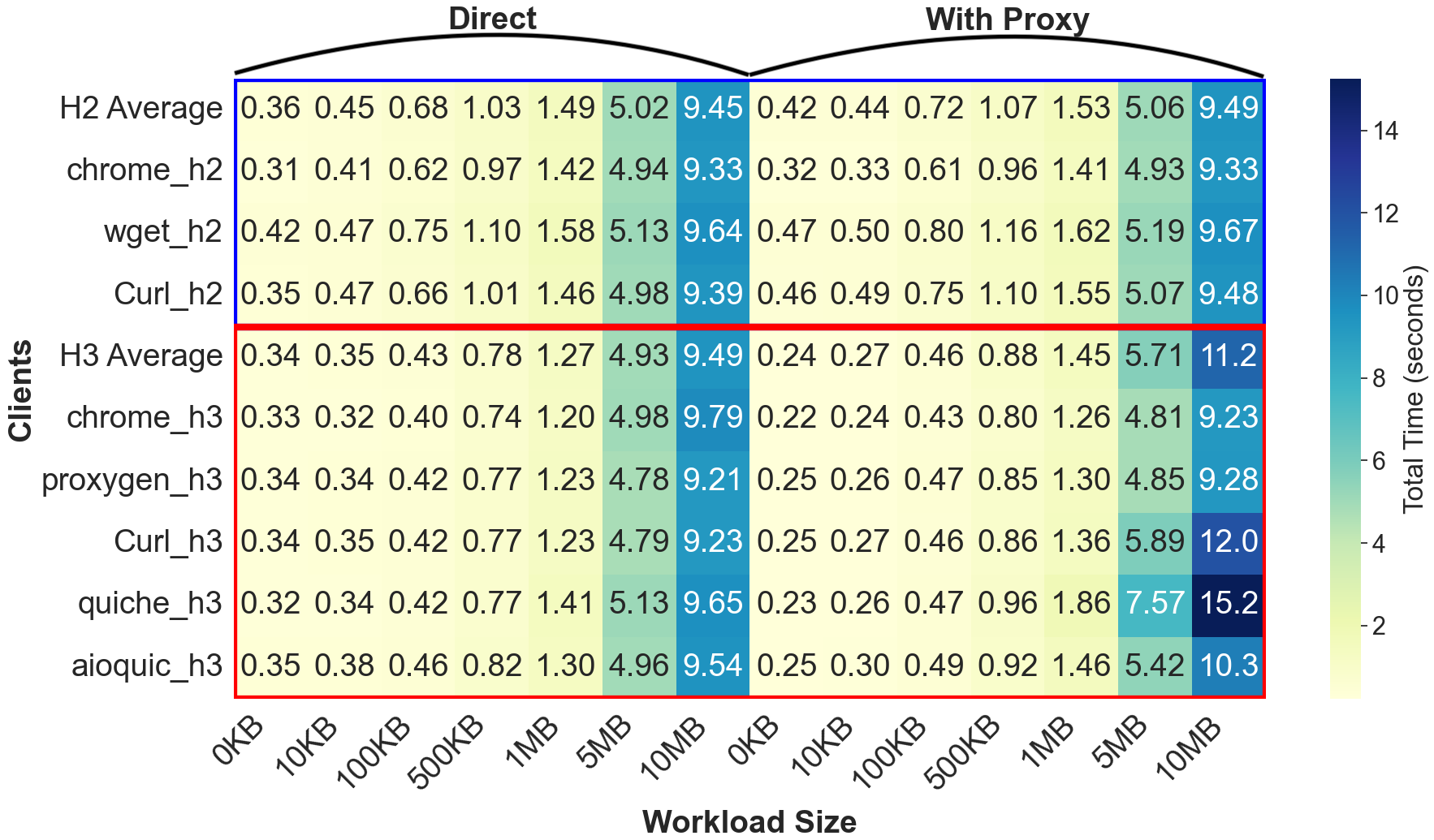}
\caption{0\% Loss, 50ms Delay}
\label{fig:1b-s}
\end{subfigure}
\hfill
\begin{subfigure}[b]{0.32\textwidth}
\centering
\includegraphics[width=\linewidth]{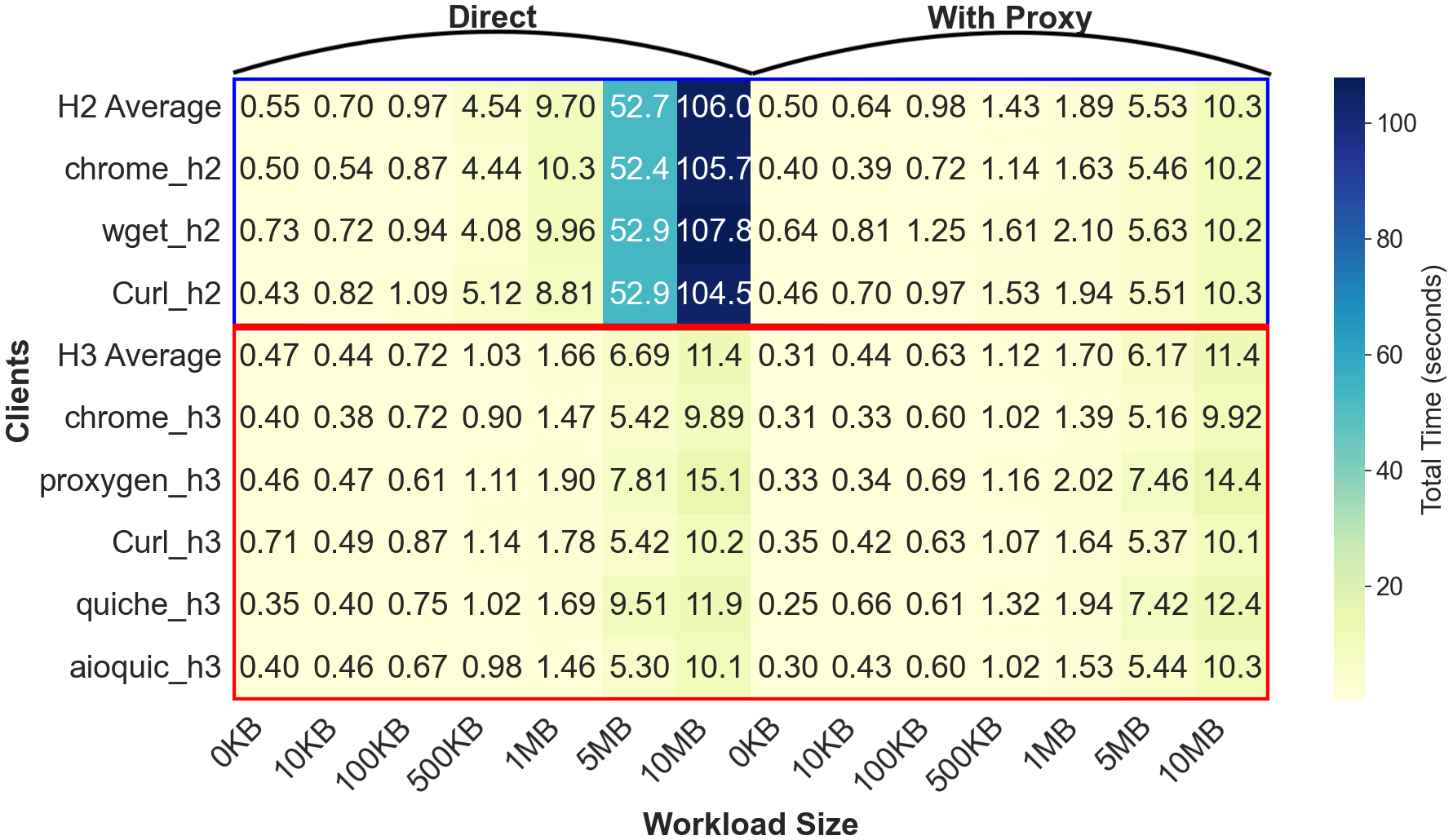}
\caption{4\% Loss, 50ms Delay}
\label{fig:1c-s}
\end{subfigure}
\caption{Single-stream results under varying loss and delay conditions. Times represent mean values over 10 runs.}
\label{fig:single_result}
\end{figure*}

\begin{figure}[htb]
  \centering
  \includegraphics[width=\linewidth, height=0.15\textheight]{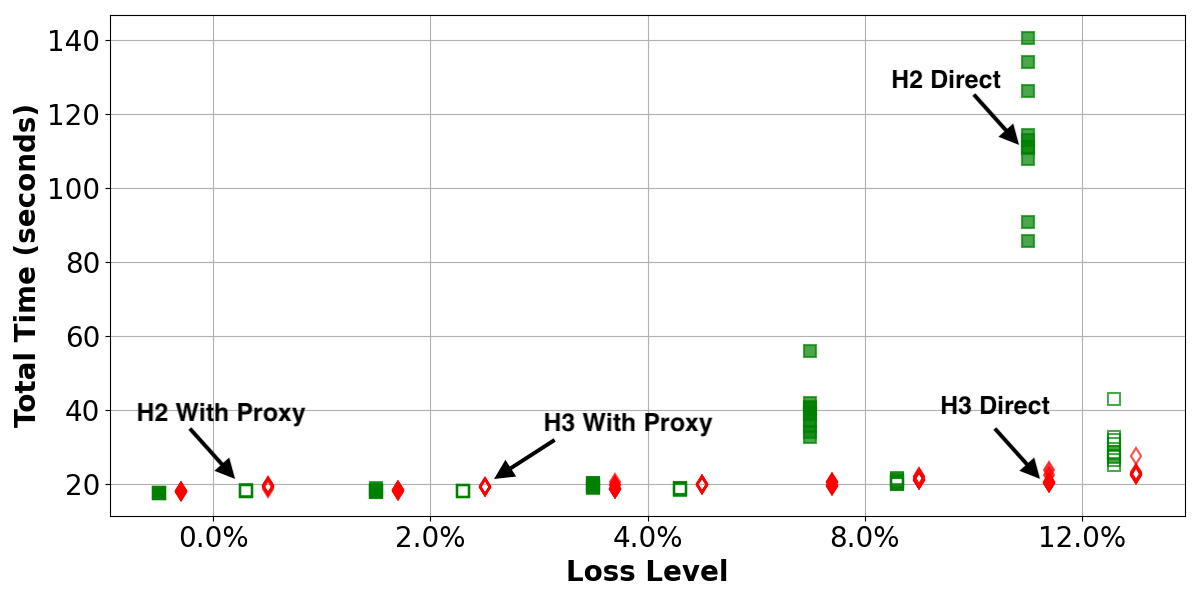}
  \caption{Multi-Stream Results. Each point represents the highest Total Request Time among 20 test files for each run. A total of 10 runs were conducted for each loss level.}
  \label{fig:multi_res}
\end{figure}

\section{Experimental Evaluation}
In this section, we present the results of our experiments and our analysis of the results.

\subsection{Single-Stream Experiment Evaluation}
\label{subsec:Single-Stream_Experiment_Evaluation}
Figure \ref{fig:single_result} presents results from single-stream experiments across various network conditions, highlighting performance differences between H2 and H3.

\textbf{Ideal Network Conditions}  
Under ideal conditions (10 Mbps, 0\% loss, 0 ms delay), H2 and H3 showed similar performance, each taking about 9 seconds to download a 10 MB file (Figure \ref{fig:1a-s}). Introducing a proxy caused a minor delay (0–1 seconds), slightly more noticeable for H3 due to Envoy's experimental H3 support compared to its optimized H2 support.

\textbf{Moderate Network Impairments}  
With moderate impairments (0\% loss, 50 ms delay), H2 and H3 remained comparable without proxies (Figure \ref{fig:1b-s}). When a proxy was introduced, H3 performance varied slightly across clients like curl and quiche, while H2 was stable, reflecting its implementation maturity. In packet loss-only scenarios (2–4\%), both protocols performed similarly.

\textbf{Severe Network Impairments}  
Under 4\% loss and 50 ms delay (Figure \ref{fig:1c-s}), performance differences widened. H3 completed the download in about 15 seconds, while H2 needed around 106 seconds without a proxy. With a proxy and BBR, H2's time decreased to 10 seconds, an improvement of 90\%. However, other CCAs like Cubic and New Reno did not show such gains. This highlights H2's sensitivity to CCA choice, whereas H3 remained consistent across CCAs, reflecting QUIC’s robust internal congestion control.

\textbf{Connection Establishment}  
We used a 0KB file to evaluate connection setup. QUIC combines TCP’s 3-way handshake and TLS 1.3 into a single exchange, saving one RTT. While both protocols performed similarly under ideal conditions, H3's faster setup became beneficial under impairments. At 4\% loss and 50 ms delay, H3 clients completed the 0KB download 14.5\% faster than H2.

\textbf{Summary of Findings}  
Network conditions heavily affect both protocols, but H3 is more resilient under high loss and latency due to QUIC’s advanced congestion control and efficient loss recovery. H2’s performance varies significantly based on network state and CCA configuration.

\textbf{Impact of CCAs}  
BBR maintains high throughput in impaired networks by adapting to bandwidth and RTT, making it more effective than loss-based CCAs like Cubic. It greatly boosts H2’s performance in proxy scenarios. However, this benefit is CCA-dependent. H3, in contrast, performs robustly across all tested CCAs, offering better consistency without external tuning.

\textbf{Proxy Influence and Implementation Maturity}  
H3’s slight proxy-related overhead likely stems from Envoy’s less mature H3 support, whereas H2 benefits from optimized handling. As H3 support evolves, this gap may shrink. Even now, H3 shows stable performance across conditions, with proxy influence being minimal compared to the significant proxy-driven improvements observed with H2 when paired with BBR.

\begin{figure*}[ht]
\centering

\begin{tikzpicture}[scale=0.6]
    \draw[fill=stableIP] (0,0) rectangle (0.6,0.3);
    \node[right] at (0.8,0.15) {Stable IP};

    \begin{scope}[xshift=4.5cm]
        \draw[fill=IPChange] (0,0) rectangle (0.6,0.3);
        \node[right] at (0.8,0.15) {IP Change};
    \end{scope}

    \begin{scope}[xshift=9.0cm]
        \draw[fill=stableIP] (0,0) rectangle (0.6,0.3);
        \draw[pattern=north east lines, pattern color=white] (0,0) rectangle (0.6,0.3);
        \node[right] at (0.8,0.15) {Stable IP w/ Proxy};
    \end{scope}

    \begin{scope}[xshift=16cm]
        \draw[fill=IPChange] (0,0) rectangle (0.6,0.3);
        \draw[pattern=north east lines, pattern color=white] (0,0) rectangle (0.6,0.3);
        \node[right] at (0.8,0.15) {IP Change w/ Proxy};
    \end{scope}
\end{tikzpicture}

\vspace{1em}

\begin{subfigure}[b]{0.32\textwidth}
\centering
\includegraphics[width=\linewidth]{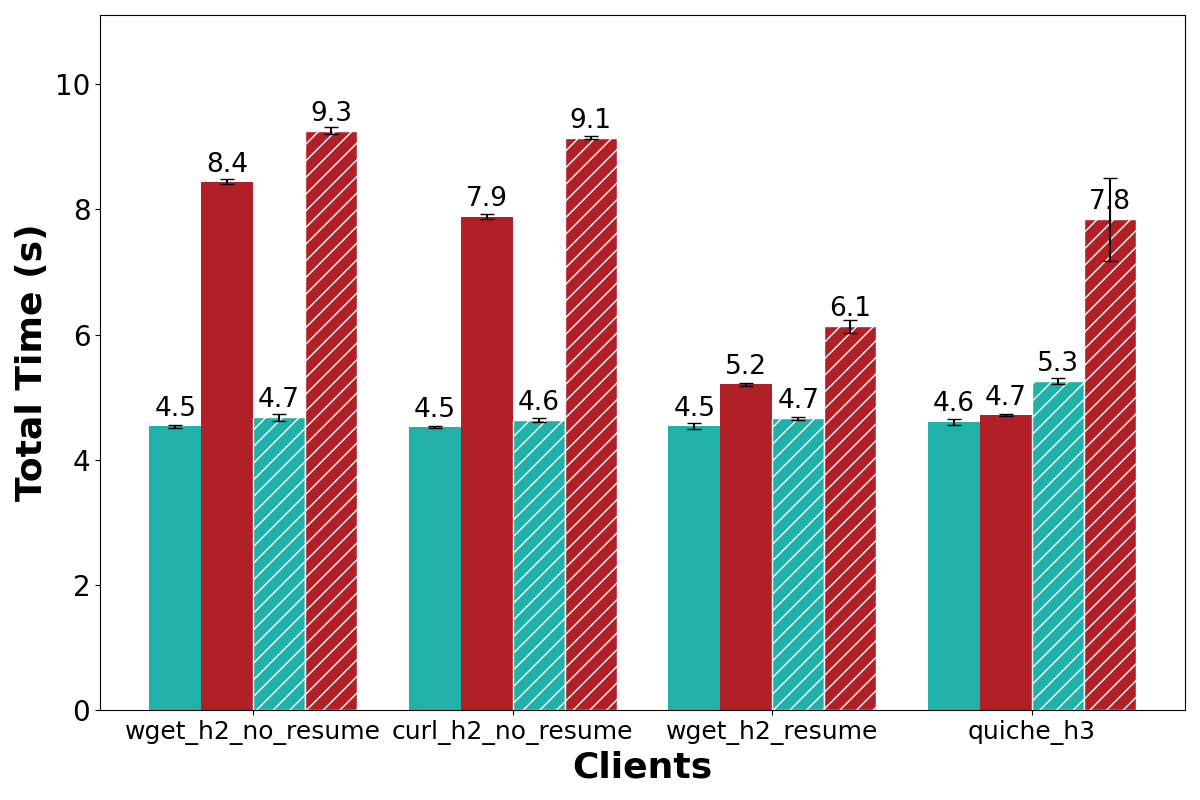}
\caption{0\% Loss, 0ms Delay}
\label{fig:1a-m}
\end{subfigure}
\hfill
\begin{subfigure}[b]{0.32\textwidth}
\centering
\includegraphics[width=\linewidth]{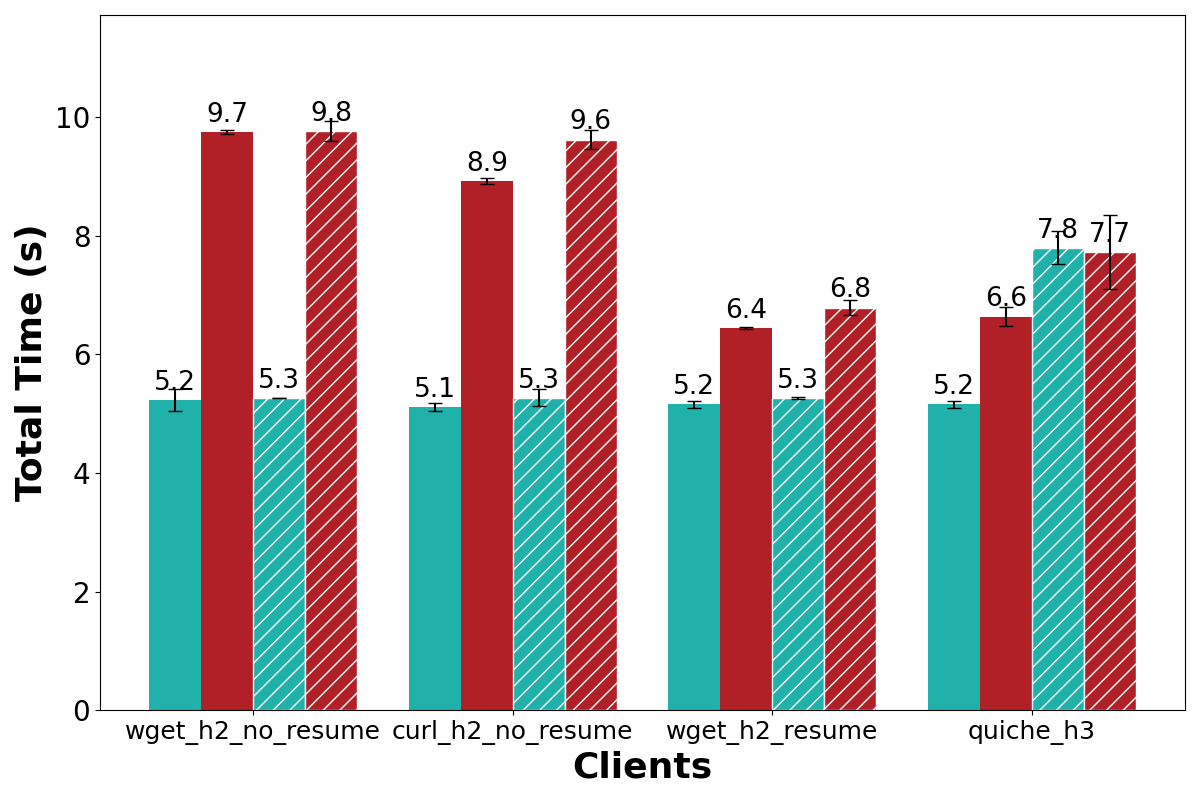}
\caption{0\% Loss, 50ms Delay}
\label{fig:1b-m}
\end{subfigure}
\hfill
\begin{subfigure}[b]{0.32\textwidth}
\centering
\includegraphics[width=\linewidth]{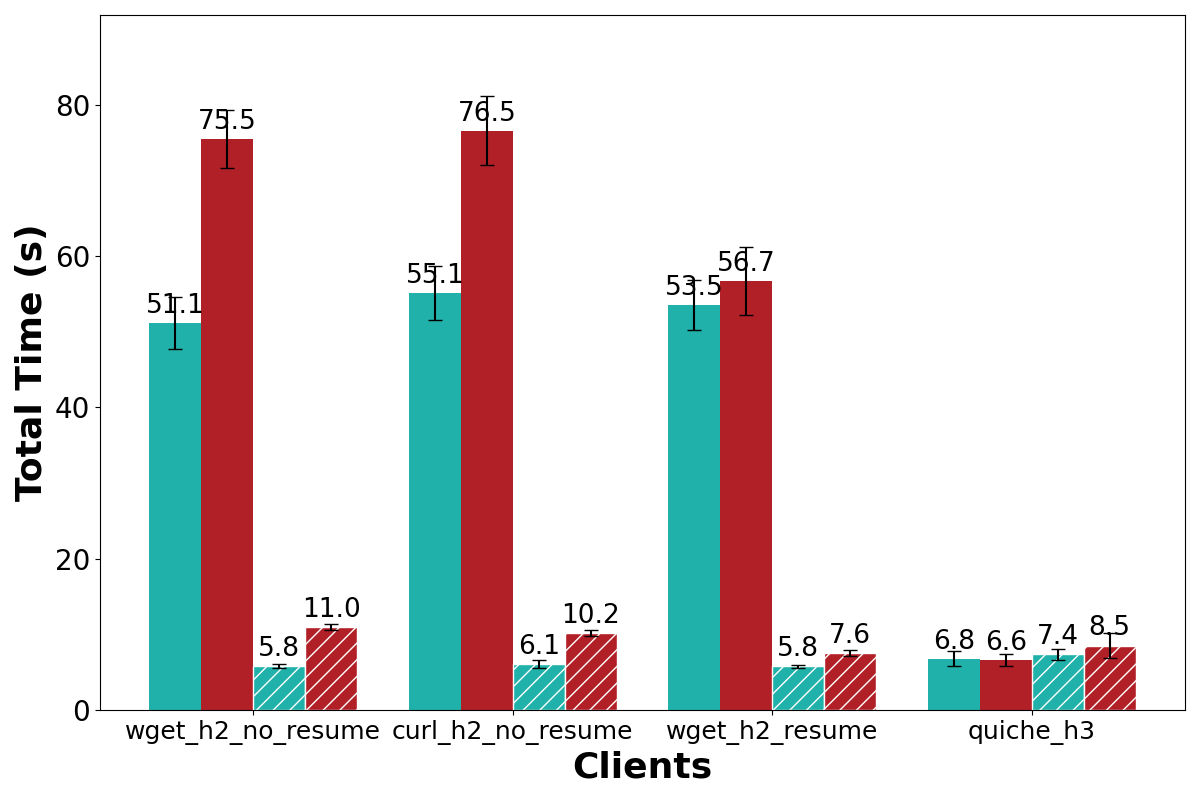}
\caption{4\% Loss, 50ms Delay}
\label{fig:1d-m}
\end{subfigure}

\caption{Connection migration results under varying network impairments. Bars show mean times over 10 runs.}
\label{fig:mig_result}
\end{figure*}

\subsection{Multi-Stream Experiment Evaluation}
Figure \ref{fig:multi_res} shows the performance of H2 and H3 under various packet loss levels using 20 concurrent streams. H3 maintains stable throughput, while H2's performance declines as loss increases. Beyond 8\% loss, H3 significantly outperforms H2, with the performance gap widening further. For example, at 12\% loss without a proxy, H2 took 113.42 seconds to download, compared to just 20.98 seconds for H3, an improvement of 81.5\%.

H3 consistently achieves stable download times when fetching 20 files simultaneously, whereas H2 exhibits high variability at higher loss levels. This inconsistency reflects H2's difficulty in maintaining performance under increased packet loss, while H3 remains robust.

H3’s resilience is largely due to QUIC’s ability to manage multiple streams within a single connection. Each QUIC stream independently controls flow and handles retransmissions, making H3 less vulnerable to stream count or packet loss. This design avoids the HoL blocking problem inherent to TCP, making H3 a more reliable option in high-loss, multi-stream scenarios.

\textbf{Analysis:} H3's key advantage in multi-streaming lies in QUIC’s architecture, which eliminates the HoL blocking common in H2. In H2, loss in one stream delays others due to TCP’s sequential delivery model, an issue that worsens with more streams or higher loss. In contrast, QUIC’s stream independence allows each stream to recover from loss independently, preserving overall throughput and stability.

QUIC's advanced congestion control, including accurate RTT tracking and adaptive loss recovery, further enhances H3’s robustness. These features allow it to quickly adjust to changing network conditions, minimizing the performance degradation seen in H2.

\textbf{Proxy Impact:} Proxies have a neutral impact on H3, performance remains stable across conditions. While H3 might show minor variation under some network impairments, the effect is minimal. In contrast, H2 benefits substantially from proxy support, especially when paired with BBR.

\subsection{Connection Migration Experiment Evaluation}
Figure \ref{fig:mig_result} shows the results of connection migration experiments, comparing H3 and H2 under varying network conditions. These results underscore QUIC's migration advantage. The H3 client (quiche\_h3) consistently achieved lower total request time during IP changes. For example, under 4\% packet loss and 50ms delay, H2 (wget\_h2\_resume) took 56.7 seconds to complete the download, whereas H3 needed only 6.6 seconds, an 88.36\% improvement.

H2 clients lacking TCP range request support (e.g., wget\_h2\_no\_resume, curl\_h2\_no\_resume) showed much higher request times and were more sensitive to network disruptions. With range requests (wget\_h2\_resume), H2’s performance improved and became comparable to H3 under either packet loss or delay alone. However, while range requests work well for static downloads, H3 is better suited for a broader set of applications, including real-time and mobile use cases.

Under high loss and delay, H3 migrations occasionally failed (in 1–2 instances), but remained rare. H2, when paired with a proxy, significantly outperformed direct H2, mainly due to the use of BBR, as discussed in Section \ref{subsec:Single-Stream_Experiment_Evaluation}.

\textbf{Analysis} H3's migration strength lies in QUIC's seamless adaptation to network path changes (e.g., switching between Wi-Fi and cellular). QUIC allows session continuity even when the client’s IP changes, minimizing delay. This benefit extends to proxy environments, where H3 remains stable despite added complexity. In contrast, H2 must re-establish a connection after an IP change, causing higher latency, especially for clients lacking TCP range requests.

H3’s ability to resume data transfer quickly makes it ideal for delay-sensitive and real-time applications. The fact that H3 rarely experiences failed migrations, even under high-loss and high-delay conditions, further underscores its robustness.

\textbf{Impact of TCP Range Requests} Range requests help H2 by enabling partial download resumption. This mitigates the absence of native migration but is better suited to static files rather than real-time data. It is a useful workaround, but not a substitute for true connection migration.

\textbf{Role of Proxies and Congestion Control} H2 benefits significantly from proxy-enhanced BBR, which improves recovery and throughput by adapting to bandwidth and delay. This makes proxies useful for boosting TCP-based protocols, albeit at the cost of added system complexity. H3, in contrast, achieves reliable performance without proxy assistance, making it preferable in scenarios requiring simplicity and robustness.

Overall, QUIC’s native support for migration gives H3 a strong edge in dynamic and unreliable networks, positioning it as a more reliable option for applications needing continuity across network transitions.

\begin{figure*}[ht]
\centering

\begin{tikzpicture}[scale=0.6]
    \draw[fill=H3] (0,0) rectangle (0.6,0.3);
    \node[right] at (0.8,0.15) {H3 Direct};

    \begin{scope}[xshift=4.2cm]
        \draw[fill=H2] (0,0) rectangle (0.6,0.3);
        \node[right] at (0.8,0.15) {H2 Direct};
    \end{scope}

    \begin{scope}[xshift=8.4cm]
        \draw[fill=H3] (0,0) rectangle (0.6,0.3);
        \draw[pattern=north east lines, pattern color=white] (0,0) rectangle (0.6,0.3);
        \node[right] at (0.8,0.15) {H3 w/ Proxy};
    \end{scope}

    \begin{scope}[xshift=13.2cm]
        \draw[fill=H2] (0,0) rectangle (0.6,0.3);
        \draw[pattern=north east lines, pattern color=white] (0,0) rectangle (0.6,0.3);
        \node[right] at (0.8,0.15) {H2 w/ Proxy};
    \end{scope}
\end{tikzpicture}

\vspace{1em}

\begin{subfigure}[b]{0.32\textwidth}
\centering
\includegraphics[width=\linewidth]{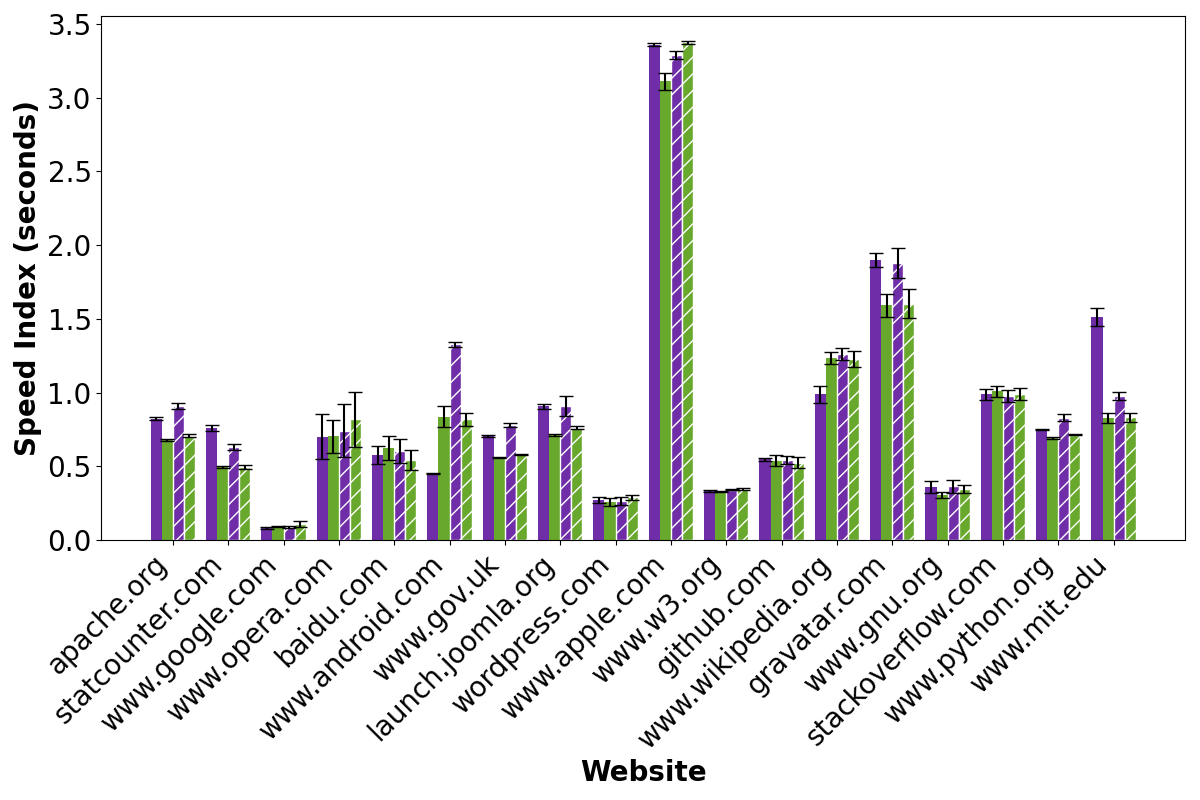}
\caption{0\% Loss, 0ms Delay}
\label{fig:1a-w}
\end{subfigure}
\hfill
\begin{subfigure}[b]{0.32\textwidth}
\centering
\includegraphics[width=\linewidth]{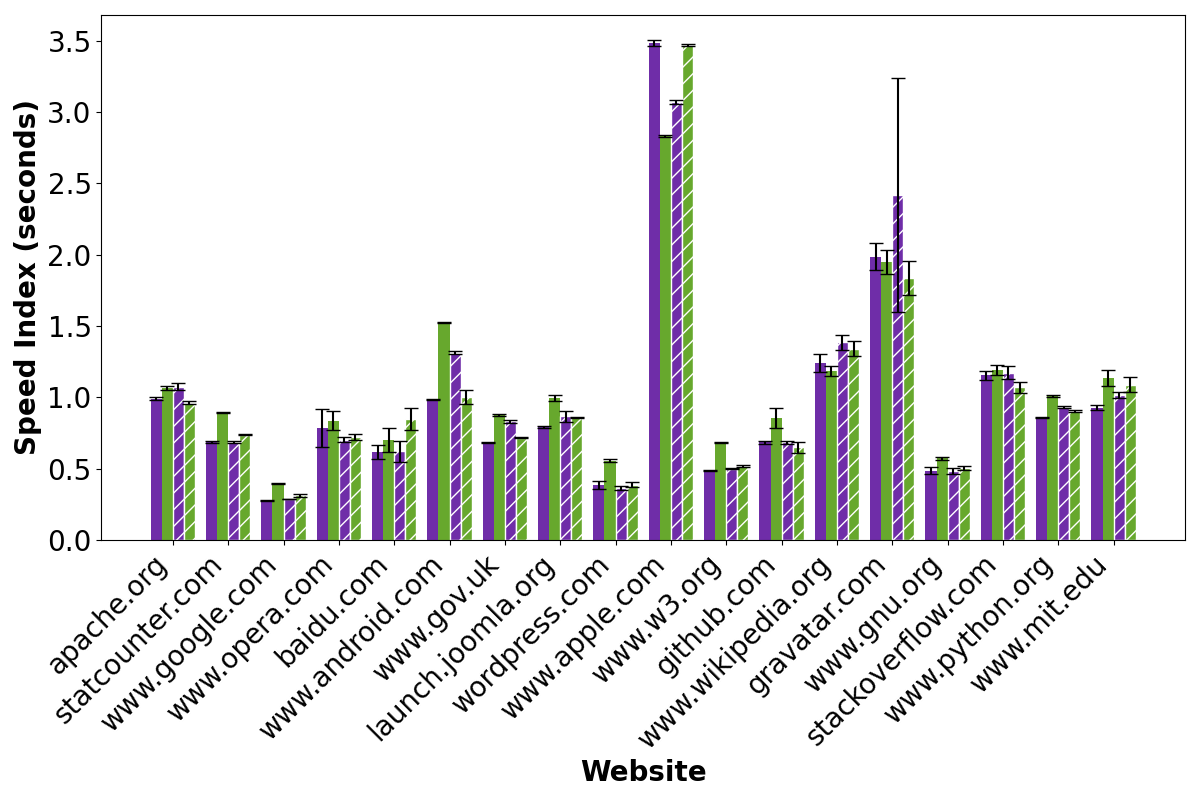}
\caption{0\% Loss, 50ms Delay}
\label{fig:1b-w}
\end{subfigure}
\hfill
\begin{subfigure}[b]{0.32\textwidth}
\centering
\includegraphics[width=\linewidth]{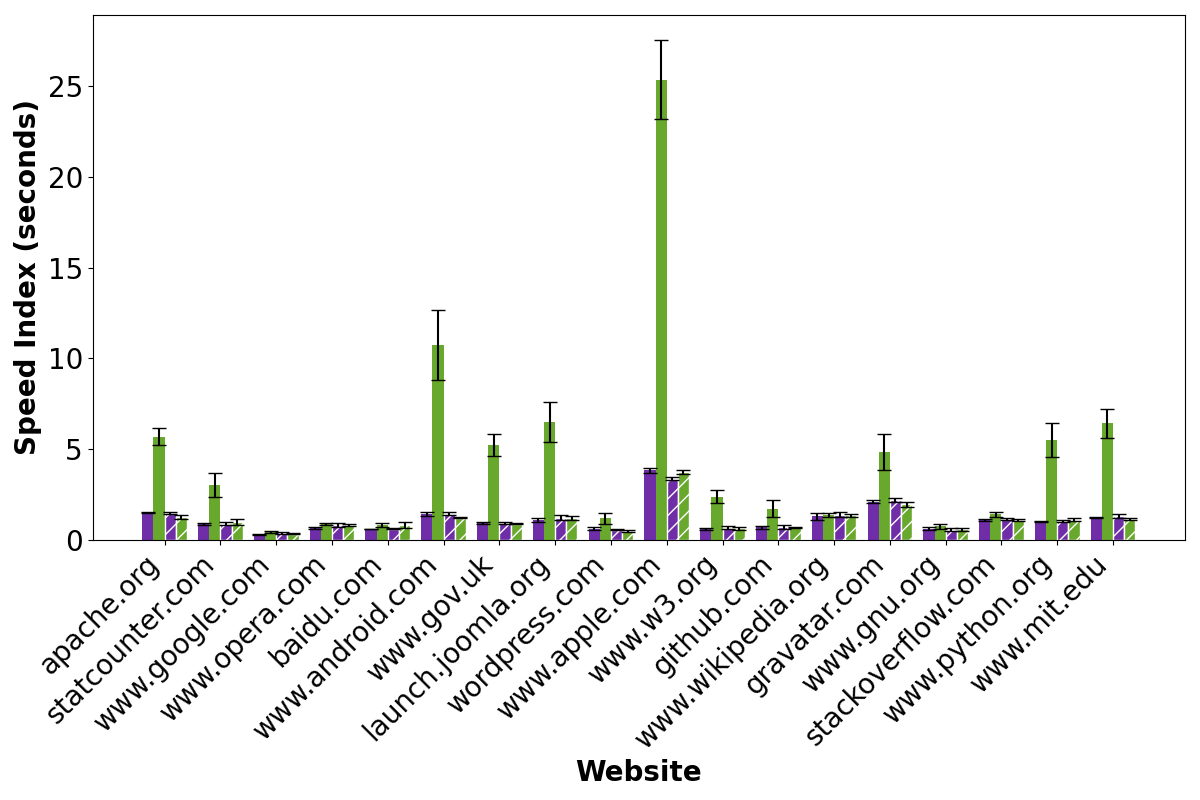}
\caption{4\% Loss, 50ms Delay}
\label{fig:1c-w}
\end{subfigure}

\caption{Web page loading results under varying network conditions. Each bar represents the mean SI over 10 runs.}
\label{fig:web_result}
\end{figure*}

\subsection{Web Page Loading Experiment Evaluation}
We evaluated the performance of H2 and H3 using SI to assess user experience during web page loading, and the results are shown in Figure \ref{fig:web_result}. SI reflects how quickly visible content appears, with scores of 0–3.4 seconds considered fast, 3.4–5.8 moderate, and over 5.8 slow.

Under baseline conditions (0\% loss, 0ms delay), shown in Figure \ref{fig:1a-w}, H2 and H3 showed comparable performance, with most websites falling in the "fast" category. Proxy usage produced mixed effects: some websites improved, others slightly degraded. Notably, more than half of the pages performed better with a proxy, even without impairments. When impairments were introduced, nearly all pages saw improved or equivalent performance using proxies.

Under 4\% packet loss and 50ms delay (Figure \ref{fig:1c-w}), H2 performance dropped considerably, with many sites rated "moderate" or "slow." In contrast, H3 remained resilient, keeping most websites in the "fast" or lower "moderate" range. H2 improved with proxy and BBR, but this gain heavily depended on the CCA. H3, by comparison, sustained strong performance without relying on external enhancements. Larger websites with more resources generally had higher SI scores, particularly under impaired conditions. Similar patterns were observed under 2\% loss and 25ms delay, but with lower SI values.

\textbf{Analysis:} The performance gap stems from QUIC’s superior handling of multiplexing. In H2, packet loss in one stream delays all others due to TCP’s sequential delivery, which hurts pages with many resources. H3 avoids this HoL blocking by independently managing streams, allowing lost packets to be retransmitted without affecting others, leading to smoother, faster page loads under poor conditions.

\textbf{Efficiency in Multiplexing:} H3 further improves bandwidth utilization by maintaining separate flow control and loss recovery for each stream. This enables parallel resource loading, which is essential for modern websites that contain many small objects. The consistently better SI for H3 in impaired conditions underscores its suitability for high-concurrency, real-world web traffic.

\textbf{Proxy Impact and Congestion Control:} Proxies with BBR help H2 by reducing latency and mitigating loss impacts. BBR’s proactive congestion control addresses TCP’s sensitivity to packet loss, lowering HoL delays. However, H3’s consistent performance with or without proxies illustrates QUIC’s ability to internally manage congestion and loss, reducing reliance on external optimizations.

Through its advanced multiplexing and independent stream control, H3 enables a more resilient and efficient browsing experience, especially in networks with variable and challenging conditions.

\subsection{Discussion}
The experimental results demonstrate H3’s superior performance over H2 across diverse network conditions, especially in scenarios involving packet loss, delay, and high concurrency. This advantage stems from QUIC’s architectural features: robust congestion control and loss recovery, efficient multiplexing that avoids HoL blocking, and seamless connection migration.

Our results show that QUIC maintains strong performance across the CUBIC, NewReno, and BBR algorithms. In contrast, HTTP/2 degrades under high loss and delay with CUBIC or NewReno. However, BBR-enabled proxies significantly enhance H2’s performance, allowing it to approach H3’s level.

We also contribute by systematically evaluating proxy impacts on QUIC’s connection migration, an area previously underexplored. Despite proxies necessitating connection termination and re-initiation, H3 performance remains stable, thanks to QUIC’s internal multiplexing and congestion control features.

These insights offer practical implications for CDN operators and developers. H3’s consistent performance across conditions suggests minimal tuning is needed, reducing operational complexity for proxy-enhanced deployments. In contrast, H2’s reliance on careful CCA selection (e.g., BBR) requires more configuration to achieve comparable performance.

While our study provides actionable guidance, several challenges remain. We do not evaluate multi-hop proxies or load balancers common in large-scale CDNs. Future research should assess H3’s behavior in edge computing environments with many concurrent users. Long-term stability of QUIC’s connection migration in evolving environments, such as 5G and cloud routing, also warrants further investigation.

\section{Conclusion and Future Work}
This study systematically evaluated the performance of H3 and H2 in proxy-enhanced environments, highlighting key differences in their resilience to network impairments, connection migration, and dependence on CCAs. Our findings show that H3 consistently maintains stable performance across diverse conditions, making it a more robust choice for modern web communication. In contrast, H2’s performance varies significantly with the chosen CCA. H3’s ability to mitigate packet loss and latency, along with seamless connection migration, makes it well-suited for mobile and high-latency environments. While H2 can achieve competitive performance with proxy-assisted enhancements, its reliance on congestion control tuning limits its adaptability.

Though this study offers valuable insights, several open challenges remain. Future work should explore QUIC’s behavior in multi-hop proxy architectures and distributed environments. Evaluating H2 and H3 under dynamic bandwidth conditions could further clarify their adaptability. Investigating AI-driven congestion control and QUIC’s migration stability in scenarios like 5G handovers and cloud routing may also yield useful advances. Additionally, our empirical findings could serve as valuable training insights for machine learning models that automatically select optimal protocols and configure network parameters based on real-time conditions.

\section*{Acknowledgment}

This work was supported by NSF CNS Award 2213672.


\begin{thebibliography}{00}

\bibitem{quicietc_2022} J. Iyengar and M. Thomson, "QUIC: A UDP-based multiplexed and secure transport," IETF, RFC 9000, May 2021.
\bibitem{http3ietc_2022} M. Bishop, "HTTP/3," IETF, RFC 9114, June 2022.
\bibitem{Jaeger2023QUICOT} B. Jaeger et al., "Quic on the highway: Evaluating performance on high-rate links," in \textit{Proc. IFIP Netw.}, 2023, pp. 1--9.
\bibitem{Liu2023SecurityAP} F. Liu and P. Crowley, "Security and performance characteristics of QUIC and HTTP/3," in \textit{Proc. 10th ACM Conf. Information-Centric Networking}, 2023, pp. 124--126.
\bibitem{Kosek2022ExploringPQ} M. Kosek, H. L. Cech, V. Bajpai, and J. Ott, "Exploring proxying QUIC and HTTP/3 for satellite communication," in \textit{Proc. 2022 IFIP Networking Conf.}, Catania, Italy, 2022, pp. 1--9.
\bibitem{border2020evaluating} J. Border, B. Shah, C.-J. Su, and R. Torres, "Evaluating QUIC's performance against performance-enhancing proxy over satellite link," in \textit{Proc. 2020 IFIP Networking Conf.}, Paris, France, 2020, pp. 755--760.
\bibitem{kosek2023secure} M. Kosek, B. Spies, and J. Ott, "Secure middlebox-assisted QUIC," in \textit{Proc. 2023 IFIP Networking Conf.}, Barcelona, Spain, 2023, pp. 1--9.
\bibitem{kuhlewind2021evaluation} M. Kühlewind, M. Carlander-Reuterfelt, M. Ihlar, and M. Westerlund, "Evaluation of QUIC-based MASQUE proxying," in \textit{Proc. 2021 Workshop Evolution, Performance, Interoperability of QUIC}, Virtual Event, 2021, pp. 29--34.
\bibitem{Carlucci2015HTTPOU} G. Carlucci, L. De Cicco, and S. Mascolo, "HTTP over UDP: An experimental investigation of QUIC," in \textit{Proc. 30th Annu. ACM Symp. Applied Computing}, Salamanca, Spain, 2015, pp. 609--614.
\bibitem{Cunha2023} B. V. Da Cunha, X. Li, W. Wilson, and K. Harfoush, "Performance benchmarking of the QUIC transport protocol," in \textit{Proc. 2023 IEEE 20th Consumer Communications Networking Conf.}, Las Vegas, NV, USA, 2023, pp. 206--212.
\bibitem{Marx2019ResourceMA} R. Marx et al., "Resource Multiplexing and Prioritization in HTTP/2 over TCP Versus HTTP/3 over QUIC," in \textit{Proc. WEBIST}, 2019, pp. 96--126.
\bibitem{Sander2022AnalyzingTI} C. Sander, I. Kunze, and K. Wehrle, "Analyzing the influence of resource prioritization on HTTP/3 HOL blocking and performance," in \textit{Proc. Int. Conf. Traffic Monitoring Analysis}.
\bibitem{Zhang2024QUIC} X. Zhang, S. Jin, Y. He, A. Hassan, Z. M. Mao, F. Qian, and Z.-L. Zhang, "QUIC is not quick enough over fast internet," in \textit{Proc. ACM Web Conf. (WWW)}, Singapore, 2024, pp. 2713--2722.
\bibitem{Shreedhar2022EvaluatingQP} T. Shreedhar et al., "Evaluating QUIC performance over web, cloud storage, and video workloads," \textit{IEEE Trans. Netw. Serv. Manag.}, vol. 19, no. 2, pp. 1366--1381, 2021.
\bibitem{wiseman2008remotely} C. Wiseman et al., "A remotely accessible network processor-based router for network experimentation," in \textit{Proc. 4th ACM/IEEE Symp. Architectures Networking Communications Systems}, San Jose, CA, USA, 2008, pp. 20--29.
\bibitem{Yu2021DissectingPO} A. Yu and T. A. Benson, "Dissecting performance of production QUIC," in \textit{Proc. Web Conf. (WWW)}, Ljubljana, Slovenia, 2021, pp. 1157--1168.
\bibitem{Tan2020ProactiveCM} L. Tan, W. Su, Y. Liu, X. Gao, N. Li, and W. Zhang, "Proactive connection migration in QUIC," in \textit{Proc. 17th EAI Int. Conf. Mobile Ubiquitous Syst.: Comput., Netw. Services}, 2020, pp. 476--481.
\bibitem{Wolsing2019APP} K. Wolsing, J. Rüth, K. Wehrle, and O. Hohlfeld, "A performance perspective on web optimized protocol stacks: TCP+TLS+HTTP/2 vs. QUIC," in \textit{Proc. Applied Networking Research Workshop}, Montreal, QC, Canada, 2019, pp. 1--7.
\bibitem{Coninck2017MultipathQD} Q. De Coninck and O. Bonaventure, "Multipath QUIC: Design and evaluation," in \textit{Proc. 13th Int. Conf. Emerging Networking Experiments Technologies}, Incheon, Republic of Korea, 2017, pp. 160--166.
\bibitem{Paiva2023AFL} T. W. do Prado Paiva, S. Ferlin, A. Brunstrom, O. Alay, and B. Y. L. Kimura, "A first look at adaptive video streaming over multipath QUIC with shared bottleneck detection," in \textit{Proc. 14th ACM Multimedia Systems Conf.}, 2023, pp. 161--172.

\end{thebibliography}
\end{document}